\documentclass[nofootinbib,a4paper,12pt,superscriptaddress,onecolumn,eqsecnum]{revtex4-1}
\usepackage{amsmath,amsfonts}
\usepackage{booktabs}
\usepackage[utf8]{inputenc}
\usepackage{siunitx} 
\usepackage{times}
\usepackage{graphicx}
\usepackage{color}
\usepackage{braket}
\usepackage{dcolumn}
\usepackage{bm,url}
\usepackage{subfigure}
\usepackage[usenames,dvipsnames,svgnames]{xcolor}  
\usepackage{hyperref}   
\hypersetup{colorlinks=true, linkcolor=blue, citecolor=red}

\definecolor{oxfordblue}{rgb}{0.0, 0.13, 0.28}
\definecolor{burgundy}{rgb}{0.5, 0.0, 0.13}
\definecolor{darkolivegreen}{rgb}{0.33, 0.42, 0.18}
\definecolor{darkblue}{rgb}{0,0,0.5}
\definecolor{richcarmine}{rgb}{0.84, 0.0, 0.25}
\definecolor{darkblue}{rgb}{0,0,0.5}
\definecolor{bluer}{rgb}{0.00,0.50,0.75}{}
\hypersetup{colorlinks=true, citecolor=red, linkcolor=blue,
	urlcolor = magenta, filecolor=magenta}

\begin{document}
	
	\newcommand\be{\begin{equation}}
		\newcommand\ee{\end{equation}}
	\newcommand\bea{\begin{eqnarray}}
		\newcommand\eea{\end{eqnarray}}
	\newcommand\bseq{\begin{subequations}} 
		\newcommand\eseq{\end{subequations}}
	\newcommand\bcas{\begin{cases}}
		\newcommand\ecas{\end{cases}}
	\newcommand{\p}{\partial}
	\newcommand{\f}{\frac}

	\title{Constraining the Tilted Dipole Cosmology with 
		Primordial Nucleosynthesis and Baryogenesis}
	\author{\textbf{Mohsen Khodadi}}
	\email{khodadi@kntu.ac.ir; mohsen.khodadi@gmail.com}
		\affiliation{School of Physics, Institute for Research in Fundamental Sciences (IPM),	P. O. Box 19395-5531, Tehran, Iran}
	\affiliation{School of Physics, Damghan University, Damghan 3671641167, Iran}
	\affiliation{Center for Theoretical Physics, Khazar University, 41 Mehseti Str., AZ1096 Baku, Azerbaijan}

\begin{abstract}
	The tilted dipole cosmology extends the standard model by incorporating a preferred spatial direction and separate bulk velocities for matter and radiation, offering a potential explanation for observed large-scale bulk flows and the CMB dipole. We conduct a comprehensive analysis of this anisotropic framework using Big Bang Nucleosynthesis (BBN) and Gravitational Baryogenesis (GB) to impose stringent constraints on the radiation tilt parameter $\beta_r$, which quantifies the magnitude of the radiation bulk flow. By deriving the modified expansion rate $H(T)$ and its impact on light element abundances, we find that the combined $2\sigma$ BBN limits from primordial helium-4 and deuterium are $|\beta_r| \lesssim 0.03$ for maximal shear (tightest bound) and $|\beta_r| \lesssim 0.2$ for minimal shear (weakest bound). These bounds are consistent with, but tighter than, those inferred from the effective neutrino species count $|\Delta N_{\mathrm{eff}}| \lesssim 0.4$. Importantly, both bounds are upper limits; $\beta_r = 0$ is always allowed by the BBN constraints. The lithium-7 problem persists, as the $\beta_r$ values required to resolve it are excluded by helium-4 data. Furthermore, GB---operating at decoupling temperatures $T_D \gtrsim 10^{12}$ GeV under the scaling $\beta_r \propto T$---yields far more severe constraints, limiting $|\beta_r^{\mathrm{BBN}}| \lesssim 10^{-8}$ for such high-scale baryogenesis scenarios.
\vspace{0.5cm}

\end{abstract}
\maketitle
\section{Introduction}
The \(\Lambda\) Cold Dark Matter (\(\Lambda\)CDM) cosmological framework, grounded in the spatially homogeneous and isotropic Friedmann--Lema\^itre--Robertson--Walker (FLRW) metric, has demonstrated considerable efficacy in characterizing both the large-scale architecture and temporal development of the Universe~\cite{Planck:2018vyg,Planck:2020olo}. This model's theoretical forecasts are in strong agreement with diverse empirical datasets, notably the anisotropies observed in the Cosmic Microwave Background (CMB), the clustering patterns of galaxies on grand scales, and the accelerated cosmic expansion indicated by Type Ia supernova luminosity-distance measurements.
Nevertheless, in spite of its observational successes, a number of enduring anomalies call into question the validity of perfect isotropy at the largest scales~\cite{Perivolaropoulos:2021jda,DiValentino:2021izs,Abdalla:2022yfr}. Specifically, galaxy surveys have detected coherent bulk motions that surpass the velocities predicted within the conventional \(\Lambda\)CDM paradigm~\cite{Watkins:2008hf,Colin:2019opb}. Moreover, the CMB dipole—typically ascribed to the Doppler effect of our own motion—displays characteristics suggestive of a possible primordial contribution~\cite{Schwarz:2015cma,Secrest:2020has}. Standard cosmology interprets this dipole almost entirely as a kinematic signal produced by the Solar System's peculiar velocity (on the order of 370 km/s with respect to the CMB frame). Yet this widely accepted explanation prompts a critical inquiry: could a portion of the dipole moment stem from inherent large-scale anisotropy rather than purely local kinematics? Such a prospect carries significant theoretical weight, as the detection of an intrinsic dipole would constitute a clear violation of statistical isotropy at the most extreme observable distances, thereby implying novel physical mechanisms extending beyond the established cosmological model~\cite{Ma:2010ps}.

Such observational indications warrant theoretical exploration of cosmological scenarios that deviate from the strict FLRW metric by incorporating either a privileged spatial direction or an intrinsic large-scale anisotropy\footnote{Modern cosmology is fundamentally grounded in the Cosmological Principle, which asserts that the Universe is both homogeneous and isotropic when viewed on sufficiently large scales. This postulate, deeply rooted in the Copernican worldview, provides the foundation for the FLRW metric.}~\cite{Krishnan:2021jmh,Aluri:2022hzs}. 

Before the inflationary paradigm became a central tenet of standard cosmology, the observed large-scale homogeneity and isotropy were widely considered to represent exceedingly fine-tuned initial conditions~\cite{Collins:1972tf}. Inflation offers a compelling dynamical resolution to this puzzle. In particular, Wald's cosmic no-hair theorem~\cite{Wald:1983ky} demonstrates that isotropy emerges as a generic late-time attractor in universes containing a positive cosmological constant, provided homogeneity is assumed. 

A clarification regarding the applicability of Wald's theorem to inflation is warranted. Most inflationary models do not strictly satisfy the assumptions underlying the theorem, and thus the theorem does not directly govern their behavior. Nonetheless, extensive research has established that inflation typically drives the Universe toward isotropy, even if complete isotropization is not guaranteed~\cite{Maleknejad:2012as}. 

Furthermore, it has been shown that a primordial dipole anisotropy could persist in superhorizon perturbation modes generated during the inflationary epoch~\cite{Turner:1991dn}. If such a mode later re-enters the Hubble horizon, it would naturally manifest as a dipole signature in cosmological observables, such as the distribution of galaxy number counts~\cite{Domenech:2022mvt}.

The possibility that the Universe may exhibit a preferred spatial direction has been examined within a range of theoretical frameworks. Notably, observations of large-scale bulk flows have revealed coherent streaming motions of galaxies and clusters that exceed the predictions of the \(\Lambda\)CDM model, with some analyses reporting tensions at the \(2.5\sigma\) level or higher~\cite{Ma:2010ps}. When combined with the observed discrepancy between the rest frame defined by the Cosmic Microwave Background and that defined by the matter distribution, these peculiar velocity measurements point to the potential existence of an intrinsic ``tilt'' in the Universe---a relative velocity between cosmic reference frames that cannot be removed by a simple Lorentz boost. Such a tilt would indicate that our motion is not merely superimposed on an otherwise homogeneous background; rather, it would suggest that the background itself carries a persistent directional signature.

Drawing on these empirical indications and the foundational work on tilted Bianchi cosmologies by King and Ellis, dipole cosmology emerges as the maximally Copernican extension of the FLRW paradigm~\cite{Ellis:1968vb,King:1972td}. The underlying rationale is conceptually straightforward: postulating only the expansion of the Universe as a foundational prior yields the standard Cosmological Principle, characterized by complete spatial isotropy. By contrast, if one additionally incorporates the observational prior that a CMB dipole exists---without presuming its entirety arises from our peculiar motion---the most symmetric scenario accommodating a bulk fluid flow along some direction reduces the isotropy group to axial symmetry, specifically Local Rotational Symmetry (LRS). This ``dipole Cosmological Principle'' maintains homogeneity while relinquishing full rotational invariance in favor of a privileged axis, thereby reviving a particular class of tilted Bianchi models (types V and VII$_h$) that have remained largely unexplored in phenomenological cosmology.

A key development rendering dipole cosmology suitable for realistic model construction is the incorporation of multiple fluid components endowed with independent bulk velocities~\cite{Ebrahimian:2023svi,Krishnan:2022qbv,Allahyari:2023kfm,Martin:2025ywz}. As underscored in these studies, conventional FLRW cosmology routinely incorporates mixtures of distinct fluids---radiation, matter (both baryonic and dark), and a cosmological constant---each governed by its own equation of state. To formulate a dipole $\Lambda$CDM model, one must consequently permit radiation and matter to carry independent tilt parameters, or rapidities, conventionally denoted $\beta_r$ and $\beta_m$, while the cosmological constant remains unaffected by the tilt. This multi-fluid generalization constitutes an essential step in transforming tilted Bianchi cosmologies from abstract dynamical systems analyses into a concrete framework equipped to confront observational tensions.

A particularly striking outcome of this theoretical framework concerns the dynamical evolution of the relative velocity between radiation and matter. Contrary to the conventional expectation that peculiar velocities should diminish as the Universe expands, the dipole $\Lambda$CDM model reveals a generic instability in which the differential tilt $\beta_r - \beta_m$ may amplify at late times across a wide range of initial configurations. This behavior---which we designate the ``tilt instability'' inherent to the Cosmological Principle---carries substantial observational implications~\cite{Krishnan:2022uar}. It implies that even a Universe originating in a near-FLRW state, possessing only a modest initial tilt in the radiation component, can experience progressive growth of anisotropic bulk flows over cosmic history, thereby potentially contributing to the observed CMB dipole and mitigating tensions manifest in late-time cosmological data.

Within this dynamical picture, the sign of the tilt parameter assumes critical importance. Although early investigations of tilted cosmologies predominantly focused on positive values of $\beta$, the dipole $\Lambda$CDM framework demonstrates that negative initial values for $\beta_r$ can generate qualitatively distinct evolutionary trajectories, including the emergence of a late-time attractor characterized by non-zero tilt. This marked sensitivity to sign underscores the fundamentally physical character of the tilt---it represents not a mere coordinate artifact but rather a genuine dynamical degree of freedom, whose observational signatures cannot be eliminated through gauge choices.

The inclusion of such tilted fluids carries significant dynamical implications. Most importantly, they serve as a source term for sustained shear anisotropy, denoted \(\sigma_{\mu\nu}\), which alters the conventional Friedmann equations describing cosmic expansion. The resulting Hubble parameter \(H(t)\) receives modifications proportional to both \(\beta^2\) and the shear amplitude, thereby changing the Universe's cooling rate. This alteration subsequently affects critical phases of thermal evolution---including Big Bang Nucleosynthesis (BBN), recombination, and CMB decoupling---by modifying freeze-out temperatures and interaction rates. As a result, even a relatively small tilt can leave detectable imprints on both the primordial abundances of light elements and the temperature anisotropy power spectrum of the CMB.

From a phenomenological perspective, this model category offers a natural and economical interpretation of several large-scale observational anomalies. The coherent bulk flows identified in galaxy surveys, together with enduring questions regarding the CMB dipole, may plausibly originate from such a primordial configuration involving cosmologically tilted fluids. In this context, tilted anisotropic cosmologies constitute an essential theoretical laboratory. They enable a systematic, parameter-driven investigation of departures from perfect isotropy. By establishing direct connections between observable dipolar features---namely the CMB dipole and large-scale bulk flows---and fundamental cosmological variables (\(\beta_m\), \(\beta_r\), \(\sigma\)), these models yield concrete and falsifiable predictions. Confronting such predictions with high-precision observational data facilitates rigorous scrutiny of the cosmic isotropy postulate and permits investigation into whether a privileged direction exists on the largest accessible scales.

Big Bang Nucleosynthesis (BBN) constitutes a powerful probe of early-Universe physics, operating within the temperature range \(T \sim 0.1\)--\(1\) MeV. The primordial abundances of light elements---specifically deuterium (D), helium-3 (\(^3\)He), helium-4 (\(^4\)He), and lithium-7 (\(^7\)Li)---exhibit remarkable sensitivity to the cosmic expansion rate during this epoch, conventionally characterized by the effective number of neutrino species \(N_{\mathrm{eff}}\)~\cite{Cyburt:2015mya,Pisanti:2007hk}. Any deviation from the standard Hubble parameter \(H(T)\), whether arising from anisotropic expansion or the presence of additional energy density components, modifies the freeze-out temperatures governing key nuclear reactions, thereby altering the final elemental yields. As such, BBN provides a precision observational laboratory for testing non-standard cosmological scenarios, including those featuring anisotropy in the early Universe~\cite{Steigman:2007xt,Iocco:2008va}.

A fundamental puzzle that persists alongside these considerations is the origin of the observed baryon asymmetry in the Universe, quantified by the baryon-to-photon ratio \(\eta_B \approx 6.1 \times 10^{-10}\). Gravitational Baryogenesis (GB)---a mechanism anticipated to operate in the pre-BBN epoch---offers an elegant pathway for generating this asymmetry via a CPT-violating interaction that couples the baryon current to the derivative of the Ricci scalar, expressed as \(\mathcal{L}_{\mathrm{GB}} \propto (\partial_\mu R) J_B^\mu\)~\cite{Davoudiasl:2004gf}. The efficiency of this mechanism hinges crucially upon the temporal evolution of \(R\), which itself depends sensitively on the underlying cosmological background. Within anisotropic frameworks such as dipole cosmology, the Ricci scalar receives additional contributions arising from both shear and spatial curvature, thereby potentially altering the magnitude of the generated baryon asymmetry and consequently imposing observational constraints on the parameter space of such models~\cite{Park:2025fmu}. In recent years, the utilization of primordial cosmological probes---notably BBN and GB---has emerged as a significant methodology for assessing the viability of extended gravity theories within the literature~\cite{Capozziello:2017bxm,Asimakis:2021yct,Khodadi:2022mzt,Lambiase:2006dq,Lambiase:2006ft,Odintsov:2016hgc,Oikonomou:2016jjh,Barrow:2020kug,Azhar:2021wvx,Troisi:2025ksj,Matei:2025epi,Sheykhi:2024fya,Sheykhi:2025zre,Luciano:2025fqg}.

This work presents a comprehensive and systematic investigation of the tilted dipole \(\Lambda\)CDM cosmology, deriving combined observational constraints from both Big Bang Nucleosynthesis (BBN) and gravitational baryogenesis (GB). We begin by deriving the modified Hubble expansion rate \(H(T, \beta_r)\) characterizing the BBN era, incorporating the effects of radiation tilt \(\beta_r\) and its associated shear anisotropy. Subsequently, we compute the resulting alterations to the primordial abundances of light elements and confront these predictions with contemporary observational measurements. In parallel, we evaluate the Ricci scalar \(R\) within the dipole cosmological framework and analyze its implications for the GB mechanism, thereby establishing upper limits on \(\beta_r\) from the requirement of reproducing the observed baryon asymmetry. Through this approach, our analysis connects the physics governing the earliest fractions of a second after the Big Bang---embodied by BBN---with potentially ultra-high-energy processes described by GB, thus furnishing a multi-epoch probe of anisotropy in the early Universe.

We structure the paper as follows: In Section \ref{dipole}, we present the governing equations of the dipole cosmology during radiation domination. Section \ref{bbn} derives the BBN constraints via the \(\Delta N_{\mathrm{eff}}\) parametrization and light element abundances. Section \ref{bg} details the calculations for GB in this framework and the resulting bounds. Finally, in Section \ref{co}, we present our conclusions and discuss the implications of our findings for anisotropic cosmological models. 

\section{Dipole Cosmology Equations during BBN}\label{dipole}
The dipole cosmology metric in comoving coordinates is \cite{Krishnan:2022qbv,Krishnan:2022uar}
\begin{equation}\label{metric}
	ds^2 = -dt^2 + a^2(t) \left[ e^{4b(t)} dz^2 + e^{-2b(t)-2A_0 z} (dx^2 + dy^2) \right],
\end{equation}
where $a(t)$ is the scale factor, $b(t)$ parametrizes anisotropy, and $A_0$ is a constant with dimensions of inverse length setting the curvature scale. The Hubble parameter and shear are defined as
\begin{equation}
	H(t) = \frac{\dot{a}}{a}, \quad \sigma(t) = 3\dot{b}.
\end{equation}
During BBN, matter density and cosmological constant are negligible, so we consider only radiation with tilt $\beta_r$. The energy-momentum tensor for tilted radiation ($p_r = \rho_r/3$) gives
\bea
&&H^2 = \frac{A_0^2}{a^2 e^{4b}} + \frac{\rho_r}{3}\left(1 + \frac{4}{3} \sinh^2\beta_r\right) + \frac{\sigma^2}{9}, \label{Hubble_eq} \\
&&\sigma = \frac{a e^{2b} \rho_r \sinh 2\beta_r}{3 A_0}, \label{shear_eq} \\
&&\dot{\rho}_r + 4H\rho_r = -4\rho_r\left(\frac{2}{3}\sigma + \coth\beta_r \, \dot{\beta}_r\right), \label{cont_eq} \\
&&(3\coth\beta_r - \tanh\beta_r)\dot{\beta}_r = -2\sigma - \frac{2\tanh\beta_r}{a}e^{-2b}. \label{beta_eq}
\eea
The conservation equation yields $\rho_r \propto (a e^{2b} \sinh\beta_r)^{-4}$.

We define dimensionless density parameters
\begin{equation}
	\tilde{\Omega}_r = \frac{\rho_r(1 + \frac{4}{3}\sinh^2\beta_r)}{3H^2}, \quad
	\tilde{\Omega}_k = \frac{A_0^2}{a^2 e^{4b} H^2}, \quad
	\tilde{\Omega}_\sigma = \frac{\sigma^2}{9H^2},
\end{equation}
satisfying $\tilde{\Omega}_r + \tilde{\Omega}_k + \tilde{\Omega}_\sigma = 1$ (see Appendix \ref{A}).

From Eq.~(\ref{shear_eq}) and $\tilde{\Omega}_k$, we find:
\begin{equation}
	\frac{\sigma}{H} = \frac{\tilde{\Omega}_r}{1 + \frac{4}{3}\sinh^2\beta_r} \cdot \frac{\sinh 2\beta_r}{\sqrt{\tilde{\Omega}_k}}.
\end{equation}
This implies:
\begin{equation}
	\tilde{\Omega}_\sigma = \frac{4 \tilde{\Omega}_r^2 \sinh^2\beta_r \cosh^2\beta_r}{(3 + 4\sinh^2\beta_r)^2 \tilde{\Omega}_k}. \label{Omega_sigma}
\end{equation}
Using the sum rule, we obtain
\begin{equation}
	H^2 = \frac{\rho_r}{3} \cdot \frac{1 + \frac{4}{3}\sinh^2\beta_r}{\tilde{\Omega}_r}. \label{H2_general}
\end{equation}
Here $\tilde{\Omega}_r$ depends on $\beta_r$ and $\tilde{\Omega}_k$ through Eqs.~(\ref{Omega_sigma}) and the sum rule.

For small tilt $|\beta_r| \ll 1$, Eq.~(\ref{Omega_sigma}) reduces to
\begin{equation}
\tilde{\Omega}_\sigma \approx \frac{4 \tilde{\Omega}_r^2 \beta_r^2}{9 \tilde{\Omega}_k}.
\end{equation}
The sum rule $\tilde{\Omega}_r + \tilde{\Omega}_k + \tilde{\Omega}_\sigma = 1$ then yields
\begin{equation}
\tilde{\Omega}_r \lesssim \frac{1}{1 + \frac{4|\beta_r|}{3}} \approx 1 - \frac{4|\beta_r|}{3},
\end{equation}
where the bound is saturated for minimal $\tilde{\Omega}_r$ (maximal shear).

Substituting into Eq.~(\ref{H2_general}) gives the effective expansion rate
\begin{equation}
H^2 \approx \frac{\rho_r}{3} \left( 1 + \frac{4}{3}\beta_r^2 \right) \left( 1 + \frac{4|\beta_r|}{3} \right) \approx \frac{\rho_r}{3} \left( 1 + \frac{4|\beta_r|}{3} + \frac{4}{3}\beta_r^2 \right). \label{H2_small_beta}
\end{equation}
Before leaving this section, a conceptual subtlety in our analysis deserves clarification: whether the radiation tilt parameter $\beta_r$ represents a genuine physical anisotropy or merely a kinematical artifact removable by observer redefinition. In a universe containing only radiation, a nonzero $\beta_r$ could indeed be absorbed by boosting to the radiation rest frame, 
rendering it kinematical. However, the dipole $\Lambda$CDM cosmology differs crucially in several respects. First, the metric itself possesses intrinsic anisotropy through the curvature parameter $A_0$ and shear $\sigma = 3\dot{b}$, which persist even in the 
radiation rest frame. Second, $\beta_r$ couples directly to these geometric quantities 
via the shear equation (\ref{shear_eq}), making it 
dynamical rather than purely kinematical. Third, even during radiation domination, 
matter is present (though subdominant), and the relative tilt $\beta_m - \beta_r$ 
between components constitutes a physical observable analogous to the CMB dipole. 
Most importantly, $\beta_r$ directly affects the expansion rate $H(T)$ through 
Eq.~(\ref{Hubble_eq}), and consequently modifies BBN abundances—observable quantities 
that cannot be altered by coordinate choices. Thus, while $\beta_r$ itself may be 
gauge-dependent, its imprints on expansion history, shear evolution, and primordial nucleosynthesis are physical, and constraints derived in the next section on $|\beta_r|$ 
represent genuine bounds on the model's anisotropic degrees of freedom.
\section{BBN Constraints through $\Delta N_{\text{eff}}$}\label{bbn}

The standard radiation energy density during BBN is given by:
\begin{equation}
	\rho_r^{\text{std}} = \frac{\pi^2}{30} g_* T^4,
\end{equation}
where $g_* \approx 10.75$ accounts for photons, electrons, positrons, and three neutrino species at $T \sim 1$ MeV. The corresponding Hubble rate in the standard FLRW cosmology is:
\begin{equation}
	H^2_{\text{std}} = \frac{\rho_r^{\text{std}}}{3}.
\end{equation}

In the dipole cosmology, the expansion rate ratio relative to the standard case can be parametrized as:
\begin{equation}
	\frac{H^2}{H^2_{\text{std}}} = 1 + \frac{7}{43} \Delta N_{\text{eff}},
\end{equation}
where $\Delta N_{\text{eff}}$ quantifies the effective change in the number of relativistic degrees of freedom due to the radiation tilt and associated anisotropy.

From Eq.~(\ref{H2_small_beta}), for small tilt $|\beta_r| \ll 1$
\begin{equation}
	\frac{H^2}{H^2_{\text{std}}} \approx 1 + \frac{4|\beta_r|}{3} + \frac{4}{3}\beta_r^2.
\end{equation}
Equating this with the $\Delta N_{\text{eff}}$ parametrization yields
\begin{equation}
\frac{4|\beta_r|}{3} + \frac{4}{3}\beta_r^2 = \frac{7}{43} \Delta N_{\text{eff}}.
\label{beta_Neff_relation}
\end{equation}

The standard model prediction for the effective number of neutrino species, accounting for non-instantaneous decoupling and finite-temperature QED effects, is $N_{\text{eff}}^{\text{SM}} = 3.044$ \cite{Mangano:2005cc, deSalas:2016ztq}. 

Current observational constraints from Planck CMB data combined with Big Bang Nucleosynthesis (primordial deuterium and helium abundances) give \cite{Planck:2018vyg, ParticleDataGroup:2022pth} (see also Refs.~\cite{Aver:2015iza,Cooke:2017cwo,Vagnozzi:2019ezj,Hsyu:2020uqb,ACT:2020gnv,Mossa:2020gjc})
\begin{equation}
	N_{\text{eff}} = 2.99 \pm 0.17 \quad (68\%\ \text{CL}).
\end{equation}
The deviation from the standard mModel prediction is therefore
\begin{equation}
\Delta N_{\text{eff}} = N_{\text{eff}} - N_{\text{eff}}^{\text{SM}} = -0.054 \pm 0.17.
\end{equation}
At $2\sigma$ confidence level ($95\%$ CL), this implies
\begin{equation}
	|\Delta N_{\text{eff}}| \lesssim 0.394 \approx 0.4.
\end{equation}
For conservative analysis, we adopt the bound $|\Delta N_{\text{eff}}| \lesssim 0.4$ throughout this work.

\begin{figure}[htbp]
	\centering
	\includegraphics[width=0.8\textwidth]{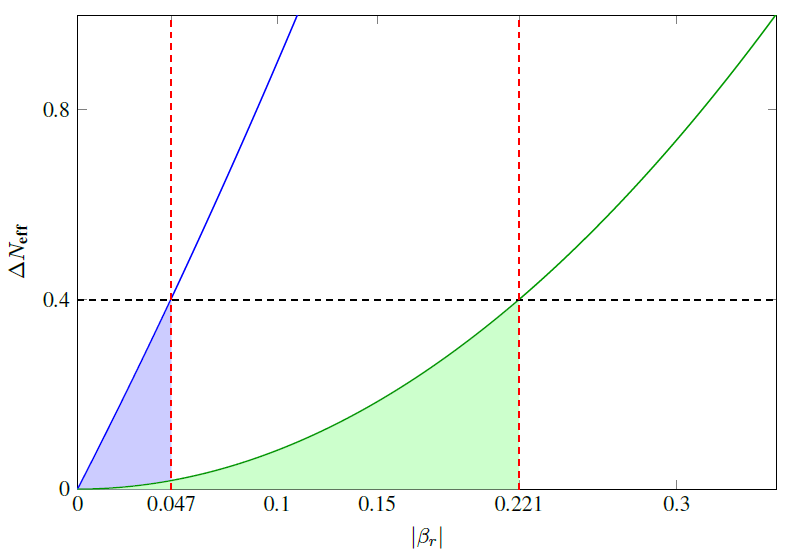}
	\caption{\textbf{BBN constraints on \(\beta_r\) from \(\Delta N_{\mathrm{eff}}\).} 
		The blue curve represents the maximal shear case (tightest bound), while the green curve corresponds to the minimal shear case (weakest bound). The shaded regions indicate the parameter space allowed by BBN constraints: \(|\beta_r| \lesssim 0.047\) for maximal shear (blue) and \(|\beta_r| \lesssim 0.221\) for minimal shear (green). The horizontal dashed line at \(\Delta N_{\mathrm{eff}} = 0.4\) indicates the \(2\sigma\) upper bound. The vertical dashed lines mark the maximum allowed values. Values of \(|\beta_r|\) greater than these bounds are excluded. Both bounds are upper limits; \(\beta_r = 0\) is always allowed.}
	\label{fig:BBN_constraints}
\end{figure}

Substituting $\Delta N_{\text{eff}} = 0.4$ into Eq.~(\ref{beta_Neff_relation}) gives
\begin{equation}
|\beta_r| + \beta_r^2 \approx 0.04884.
\end{equation}
Solving this quadratic equation for $|\beta_r|$, 
in the maximal shear case where the effect on $H(T)$ is largest, we obtain the tightest bound:
\begin{equation}
	|\beta_r| \lesssim 0.047 \quad (2\sigma\ \text{BBN bound, maximal shear}).
	\label{bound_max_shear_tight}
\end{equation}
If spatial curvature $\tilde{\Omega}_k$ is sufficiently large to suppress shear, then $\tilde{\Omega}_r \approx 1$, and Eq.~(\ref{H2_general}) reduces to $H^2/H^2_{\text{std}} \approx 1 + \frac{4}{3}\beta_r^2$. In this limit
\begin{equation}
|\beta_r| \lesssim 0.221.
\end{equation}
Therefore, the general BBN constraint on the radiation tilt parameter, accounting for the range of possible shear contributions, is:
\begin{equation}
	|\beta_r| \lesssim 
	\begin{cases}
		0.047, & \text{maximal shear (tightest bound)},\\
		0.221, & \text{minimal shear (weakest bound)},
	\end{cases}
	\quad \text{at } 2\sigma \text{ confidence}. 
\end{equation}
Both bounds are upper limits, and $|\beta_r|= 0$ is always allowed by the BBN constraints.
Figure \ref{fig:BBN_constraints} shows the allowed region in the $|\beta_r| - \Delta N_{\mathrm{eff}}$ plane. The blue shaded region corresponds to the maximal shear case ($\tilde{\Omega}_r$ minimal), where the tightest bound applies. The green shaded region corresponds to the minimal shear case, where shear is suppressed by spatial curvature, allowing larger values of $|\beta_r|$. The horizontal dashed line at $\Delta N_{\mathrm{eff}} = 0.4$ indicates the $2\sigma$ upper bound from observational constraints. The black dashed curves represent the relationship between $|\beta_r|$ and $\Delta N_{\mathrm{eff}}$ from Eq.~(\ref{beta_Neff_relation}): the upper curve (blue) corresponds to maximal shear, while the lower curve (green) corresponds to minimal shear. Points below each curve are allowed by BBN constraints, while points above would violate the $|\Delta N_{\mathrm{eff}}| < 0.4$ constraint. The vertical dashed lines mark the maximum allowed values of $|\beta_r|$: $0.047$ for maximal shear and $0.221$ for minimal shear. Importantly, both bounds are upper limits; $\beta_r = 0$ is always allowed by the BBN constraints.

These constraints have several important implications for the dipole cosmology model. First, the radiation tilt during BBN ($T \sim 1$ MeV) must be very small, satisfying $|\beta_r| \lesssim 0.047$ in the maximal shear scenario and $|\beta_r| \lesssim 0.221$ in the minimal shear scenario. Conservatively, $|\beta_r| < 0.22$ regardless of the curvature/sheaf configuration. Second, if the late-time attractor solution with negative $\beta_r$ exists, its magnitude during BBN must satisfy these bounds. Third, the decay of $\beta_r$ from BBN to the present (where CMB dipole measurements give $\beta_0 \sim 10^{-3}$) must be consistent with the tilt evolution equations of the model. Fourth, the constraints are compatible with the model's prediction that initial anisotropy is washed out at late times, as $\sigma \propto e^{-3Ht}$.

The small allowed range for $|\beta_r|$ during BBN suggests that any primordial radiation bulk flow must have been modest, with a corresponding velocity $v_r / c = \tanh \beta_r \lesssim 0.047$ (maximal shear) or $v_r / c \lesssim 0.22$ (minimal shear).

\section{BBN Constraints from Light Element Abundances}
Let us derive BBN constraints on the dipole cosmology using individual light element abundances. This is more robust than just using $\Delta N_{\text{eff}}$ since different elements have different sensitivities to the expansion rate.
The primordial abundances depend on:  Expansion rate $H(T)$  determining freeze-out times,  Baryon-to-photon ratio $\eta = n_b/n_\gamma$, and Neutron lifetime $\tau_n$.  For dipole cosmology, only $H(T)$ is modified.

The expansion rate during BBN is modified according to \footnote{Maximal shear occurs when $\tilde{\Omega}_\sigma$ is maximized for given $\beta_r$. 
Minimizing $\tilde{\Omega}_r$ subject to the sum rule yields
$
\tilde{\Omega}_r \approx \frac{1}{1 + \frac{4}{3}|\beta_r|} \approx 1 - \frac{4}{3}|\beta_r|.
$ which leads to the following Hubble rate ratio
\[
\frac{H^2}{H_{\text{std}}^2} = \frac{1 + \frac{4}{3}\beta_r^2}{\tilde{\Omega}_r} 
\approx \frac{1 + \frac{4}{3}\beta_r^2}{1 - \frac{4}{3}|\beta_r|} 
\approx 1 + \frac{4}{3}|\beta_r| + \mathcal{O}(\beta_r^2).
\]
Minimal shear occurs when $\tilde{\Omega}_\sigma \approx 0$. In this limit, $\tilde{\Omega}_r \approx 1$ and:
\[
\frac{H^2}{H_{\text{std}}^2} \approx 1 + \frac{4}{3}\beta_r^2, \\
\Rightarrow \quad H \approx H_{\text{std}} \sqrt{1 + \frac{4}{3}\beta_r^2} 
\approx H_{\text{std}} \left(1 + \frac{2}{3}\beta_r^2\right).
\]}
\begin{align}
	H(T) &= H_{\text{std}}(T) \left(1 + \frac{2|\beta_r|}{3}\right) \quad \text{(maximal shear)},\\
	H(T) &= H_{\text{std}}(T) \left(1 + \frac{2}{3}\beta_r^2\right) \quad \text{(minimal shear)},
\end{align}
for $|\beta_r| \ll 1$.

In standard BBN, the $\text{n/p}$ ratio freezes at $T_f \approx 0.8$ MeV when:
$
\Gamma_{n\leftrightarrow p}(T_f) \approx H(T_f)
$
where $\Gamma_{n\leftrightarrow p} \propto T^5$. 
In dipole cosmology:
\be
T_f \approx T_f^{\text{std}} \left[ \frac{H(T_f)}{H_{\text{std}}(T_f)} \right]^{1/3}
\ee
For maximal and minimal shears:
\bea
&&T_f^{\text{Max}} \approx T_f^{\text{std}} \left( 1 + \frac{2}{9} |\beta_r|\right)\\
&&T_f^{\text{Min}} \approx T_f^{\text{std}} \left( 1+  \frac{2}{9}\beta_r^2 \right)
\eea respectively.
The freeze-out n/p ratio:
\be
(n/p)_f = \exp\left(-\frac{\Delta m}{T_f}\right), \quad \Delta m = 1.293 \text{ MeV}
\ee
Thus for maximal and minimal shears, we have
\bea
&&\left(\frac{n}{p}\right)_f^{\text{Max}} \approx 0.2 \times \exp(1.615|\beta_r|)\\
&&\left(\frac{n}{p}\right)_f^{\text{Min}} \approx 0.2 \times \exp(0.36\beta_r^2)
\eea
where 0.2 is the standard value of $\left(\frac{n}{p}\right)_f^{\text{std}}$ \cite{Boesgaard:1985km,Cyburt:2015mya,Pitrou:2018cgg,Bernstein:1988ad}.

Now, we are in a suitable position to evaluate of individual element constraints.

\subsubsection{Helium-4 ($Y_p$)}
Primordial helium abundance is most sensitive to $\text{n/p}$ freeze-out:
\bea
Y_p \approx \frac{2(n/p)_f}{1 + (n/p)_f} &\approx 0.2470 + 0.185|\beta_r| \quad \text{(max shear)}\\
&	\approx 0.2470 + 0.124\beta_r^2 \quad \text{(min shear)}
\eea
By setting $Y_p^{\text{obs}} = 0.2453 \pm 0.0066$ \cite{Aver:2020fon}, $Y_p^{\text{obs}} = 0.2458 \pm 0.0013$ \cite{Aver:2026dxv}, we arrive at
\begin{equation}
|\beta_r| < 0.023 \quad \text{(max shear)}, \quad |\beta_r| < 0.186 \quad \text{(min shear)}.
\end{equation}

\subsubsection{Deuterium (D/H)}
The deuterium abundance is determined by the freeze-out of the 
destruction reaction $D(p,\gamma)^3$He. The freeze-out condition is
\begin{equation}
\Gamma_{D(p,\gamma)^3\text{He}}(T_f^D) = n_p \langle\sigma v\rangle_{D(p,\gamma)} \approx H(T_f^D),
\end{equation}
where the reaction rate scales as $\langle\sigma v\rangle \propto T^{-2/3}\exp(-B/T)$ with $B \approx 5.5$ MeV.

For the maximal shear case with $H(T) = H_{\text{std}}(T)(1 + 2|\beta_r|/3)$, 
solving the freeze-out condition yields
\begin{equation}
	T_f^D \approx T_f^{D,\text{std}} + 8.76\times10^{-4} |\beta_r| \ \text{MeV},
\end{equation}
where $T_f^{D,\text{std}} \approx 0.085$ MeV is the standard freeze-out temperature. The deuterium abundance at freeze-out scales as
\begin{equation}
	\frac{n_D}{n_b} \propto \exp\left(-\frac{Q_D}{T_f^D}\right),
\end{equation}
with $Q_D = 2.225$ MeV (deuteron binding energy). This gives
\begin{align}
\frac{(D/H)}{(D/H)^{\text{std}}} &\approx \exp\left[-Q_D\left(\frac{1}{T_f^D} - \frac{1}{T_f^{D,\text{std}}}\right)\right] \\
	&\approx \exp(-0.269|\beta_r|) \quad \text{(maximal shear)}.
\end{align}
For the minimal shear case ($H(T) = H_{\text{std}}(T)(1 + 2\beta_r^2/3)$)
\begin{equation}
\frac{(D/H)}{(D/H)^{\text{std}}} \approx \exp(-0.269\beta_r^2) \quad \text{(minimal shear)}.
\end{equation}
Using $(D/H)^{\text{obs}} = (2.527 \pm 0.030) \times 10^{-5}$ \cite{Cooke:2017cwo}, and  $(D/H)^{\text{obs}} = (2.51 \pm 0.11) \times 10^{-5}$ \cite{Yeh:2020mgl}, it results in the following upper bounds
\begin{equation}
|\beta_r| < 0.124 \quad \text{(max shear)}, \quad |\beta_r| < 0.352 \quad \text{(min shear)}.
\end{equation}

\subsubsection{Helium-3 ($^3$He/H)}
The faster expansion generally increases $^3$He/H because $^3$He destruction reactions freeze earlier, and the decrease in production from D is less important than the decrease in destruction. From BBN codes: $^3\text{He}/H \propto H^{0.2}$ \cite{Smith:1992yy,Kawano:1992ua}.

Let us consider scaling relation in dipole cosmology. Since $^3\text{He}/H \propto H^{0.2}$, thereby, for the maximal shear ($H \propto 1 + \frac{2|\beta_r|}{3}$):
\be
\frac{^3\text{He}}{H} \approx (^3\text{He}/H)^{\text{std}} \left(1 + \frac{2|\beta_r|}{3}\right)^{0.2}
\ee
so that for small $|\beta_r|$, leads to 
\be
(^3\text{He}/H) \approx (1.04\times10^{-5})(1 + 0.133|\beta_r|)
\ee
In the same manner for the minimal shear ($H \propto 1 + \frac{2}{3}\beta_r^2$), we have
\be
(^3\text{He}/H) \approx (1.04\times10^{-5})(1 + 0.133\beta_r^2)
\ee
Using $^3\text{He}/H=1.1 + 2\times0.2 = 1.5\times10^{-5}$ \cite{Bania:2002yj} (within $2\sigma$ limit), we arrive at the following upper bounds
\begin{equation}
|\beta_r| < 3.326 \quad \text{(max shear)}, \quad |\beta_r| < 1.824 \quad \text{(min shear)}.
\end{equation}

\subsubsection{Lithium-7 ($^7$Li/H)}
Observations of $^7$Li in metal-poor halo stars give 
$(^7\text{Li}/H)^{\text{obs}} = (1.58 \pm 0.31) \times 10^{-10}$ \cite{Sbordone:2010zi,Cyburt:2015mya}. The standard BBN prediction is $(^7\text{Li}/H)^{\text{std}} = (5.02 \pm 0.25) \times 10^{-10}$,
leading to the well-known lithium problem (factor $\sim 3$ discrepancy) \cite{Yeh:2020mgl, Fields:2022mpw}.

The primordial abundance of \(^7\)Li is primarily produced through the decay of \(^7\)Be, which is synthesized via the radiative capture reaction \(^3\)He\((\alpha,\gamma)^7\)Be, followed by electron capture \(^7\)Be\((e^-,\nu_e)^7\)Li. 

A faster cosmic expansion rate \(H(T)\) during the BBN epoch affects the final \(^7\)Li abundance through two competing mechanisms:

\textbf{Earlier freeze-out of \(^7\)Be production:} The reaction \(^3\)He\((\alpha,\gamma)^7\)Be freezes out at higher temperatures when the expansion rate is larger. This reduces the time available for beryllium synthesis, leading to a lower \(^7\)Be abundance and consequently less \(^7\)Li production.
	
\textbf{Earlier freeze-out of \(^7\)Li destruction:} The destruction channels of \(^7\)Li (such as \(^7\)Li\((p,\alpha)^4\)He) also freeze out earlier in a faster expanding Universe. This reduces the amount of \(^7\)Li that gets destroyed.

The production effect (item 1) dominates over the destruction effect (item 2). As a result, a faster expansion rate leads to a net decrease in the primordial \(^7\)Li abundance. Numerical BBN simulations yield the scaling relation: $\frac{^7\text{Li}}{H} \propto H^{-0.8}$. This power-law dependence indicates that a modest increase in the Hubble expansion rate results in a significant suppression of the final lithium yield, with an exponent of \(-0.8\).

For dipole cosmology:
\begin{align}
	\text{Maximal shear: } & \frac{^7\text{Li}}{H} \approx (5.02\times10^{-10})(1 - 0.533|\beta_r|), \\
	\text{Minimal shear: } & \frac{^7\text{Li}}{H} \approx (5.02\times10^{-10})(1 - 0.533\beta_r^2),
\end{align}
we are facing with the following  weak constraints:
\begin{align}
	\text{Maximal shear: } & |\beta_r| < 1.52 \quad (2\sigma,\ \text{lower limit}), \\
	\text{Minimal shear: } & |\beta_r| < 1.23 \quad (2\sigma,\ \text{lower limit}),
\end{align}
where are far weaker than those from $Y_p$ and $\text{D/H}$.

\subsubsection{Joint constraints}

In general, Helium-3 ($^3$He/H), and Lithium-7 ($^7$Li/H) provide very weak constraints ($|\beta_r|$). As a result, useful constraints come only from $Y_p$ and D/H.

Combining $Y_p$ and D/H abundances in a $\chi^2$ analysis i.e.,
\begin{equation}
	\chi^2 = \left(\frac{Y_p - Y_p^{\text{obs}}}{\sigma_Y^{\text{eff}}}\right)^2 
	+ \left(\frac{(D/H) - (D/H)^{\text{obs}}}{\sigma_D^{\text{eff}}}\right)^2,
\end{equation}
where $\sigma_Y^{\text{eff}} = \sqrt{\sigma_Y^2 + \sigma_{Y,\text{th}}^2}$ and 
$\sigma_D^{\text{eff}} = \sqrt{\sigma_D^2 + \sigma_{D,\text{th}}^2}$ include theoretical uncertainties, gives the tightest constraints:
\bea
&&|\beta_r| < 0.030 \quad (2\sigma,\ \text{maximal shear}),
\\
&&|\beta_r| < 0.2 \quad (2\sigma,\ \text{minimal shear}).
\eea
These are consistent with, but slightly tighter than, the $\Delta N_{\text{eff}}$ constraint of $|\beta_r|$ obtained in previously.
The $^4$He abundance provides the strongest constraint due to its direct dependence on the n/p freeze-out ratio. The fact that $^4$He and $\text{D/H}$ give consistent bounds suggests the dipole cosmology with small $|\beta_r|$ is compatible with primordial nucleosynthesis. 

\section{GB in Dipole Cosmology}\label{bg}
The GB mechanism introduces a CPT-violating coupling between the baryon current and the Ricci scalar \cite{Davoudiasl:2004gf} as follows
\begin{equation}
	\mathcal{L}_{\mathrm{GB}} = \frac{g_b}{M_\ast^2} (\partial_\mu R)J_B^\mu,
\end{equation}
where $M_{*}$ is a cutoff scale (typically $\sim M_{\mathrm{Pl}}$), $g_{b} \sim \mathcal{O}(1)$ is the effective coupling constant, and $J_{B}^{\mu}$ is the baryon current defined as
\begin{equation}
	J_{B}^{\mu} = \sum_{i} q_{i} \bar{\psi}_{i} \gamma^{\mu} \psi_{i}, 
\end{equation}
with $\psi_{i}$ and $\bar{\psi}_{i} \equiv \psi_{i}^{\dagger}\gamma^0$ denoting the Dirac spinor field and its Dirac adjoint, respectively, $\gamma^{\mu}$ are the Dirac gamma matrices satisfying the Clifford algebra $\{\gamma^{\mu}, \gamma^{\nu}\} = 2\eta^{\mu\nu}$, and $q_{i}$ is the baryon number of species $i$. This current is associated with the global $U(1)_B$ baryon number symmetry of the Standard Model.
\\
this coupling generates an effective chemical potential for baryons in thermal equilibrium
\begin{equation}
	\mu_{B} = \frac{g_b \dot{R}}{M_{*}^{2}},
\end{equation}
which creates a baryon-antibaryon asymmetry. The resulting baryon-to-entropy ratio is then given by
\begin{equation}\label{B}
	\eta_{B} \equiv \frac{n_{B}}{s} \approx \frac{15g_{b}}{4\pi^{2}g_{*s}}\frac{\dot{R}}{M_{*}^{2}T}\bigg|_{T_{D}}, \tag{5.4}
\end{equation}
where $T_{D}$ is the decoupling temperature and $g_{*s}$ is the effective number of relativistic degrees of freedom contributing to the entropy density.

\begin{figure}[htbp]
	\centering
	\includegraphics[width=0.8\textwidth]{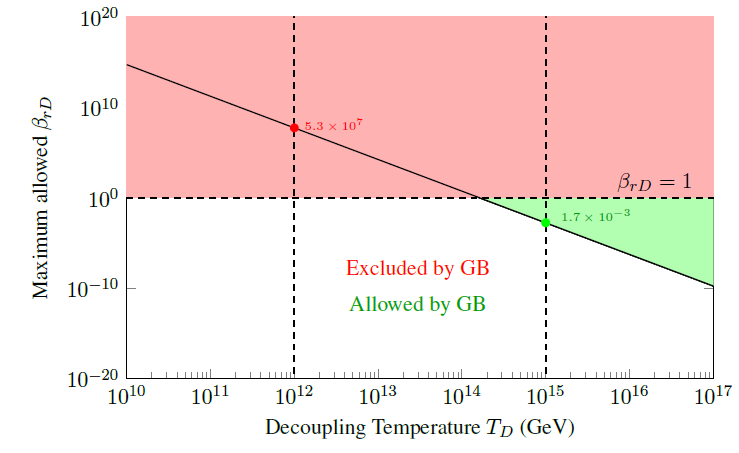}
	\caption{The allowed and excluded regions in the $T_D-\beta_{rD}^{\max}$ parameter space in light of constraint $|\eta_B^{\text{dipole}}| \lesssim \eta_B^{\text{obs}}$.}
	\label{fig:Td}
\end{figure}

We need compute the Ricci scalar for the metric (\ref{metric}). To do so, we
begin with compute the Christoffel symbols
\bea
&&\Gamma^0_{11} = a^2 e^{4b} (H + 2\dot{b}), \quad \Gamma^0_{22} = \Gamma^0_{33} = a^2 e^{-2b-2A_0 z} (H - \dot{b}),
\\ \nonumber
&&\Gamma^1_{01} = H + 2\dot{b}, \quad \Gamma^1_{22} = -A_0 a^2 e^{-4b-2A_0 z}, \quad \Gamma^1_{33} = -A_0 a^2 e^{-4b-2A_0 z},
\\ \nonumber
&&\Gamma^2_{02} = H - \dot{b}, \quad \Gamma^2_{12} = -A_0, \quad \Gamma^3_{03} = H - \dot{b}, \quad \Gamma^3_{13} = -A_0~,
\eea
and Ricci tensor components
\bea
&&R_{00} = -3\dot{H} - 3H^2 - 6\dot{b}^2,
\\ \nonumber
&&R_{11} = a^2 e^{4b} \left[ \dot{H} + 3H^2 + 2\ddot{b} + 6H\dot{b} + 2\dot{b}^2 + \frac{2A_0^2}{a^2} e^{-4b} \right],
\\ \nonumber
&&R_{22} = R_{33} = a^2 e^{-2b-2A_0 z} \left[ \dot{H} + 3H^2 - \ddot{b} - 3H\dot{b} + \dot{b}^2 + \frac{2A_0^2}{a^2} e^{-4b} \right].
\eea
After computation $R = g^{\mu\nu} R_{\mu\nu}$, we arrive at
\be \label{R}
R = 6\dot{H} + 12H^2 + 2\dot{\sigma} + 6H\sigma + 12H^2\tilde{\Omega}_\sigma + 6H^2\tilde{\Omega}_k.
\ee
During radiation domination with small tilt ($\beta_r \ll 1$) and shear 
($\sigma \ll H$), the leading correction to the standard $\dot{R} \approx 0$ is
\begin{align}
	\dot{R} &\approx -12H^4\beta_r^2 - 8H^3\beta_r\sigma - 8\frac{H^3\beta_r^3}{a}e^{-2b} \nonumber \\
	&\quad - 2H\sigma^2 + \frac{8}{3}H^2\sigma\beta_r^2 + \frac{3}{H}(\dot{H}^2\tilde{\Omega}_k).
\end{align}
Substitute $\dot{R}$ into the GB formula (\ref{B}), we have
\bea\label{dipol}
&&\eta_B^{\text{dipole}} \approx \frac{15g_b}{4\pi^2 g_*} \frac{1}{M_*^2 T_D} \times \nonumber \\
&&\left[ -12H_D^4\beta_{rD}^2 - 8H_D^3\beta_{rD}\sigma_D - 8\frac{H_D^3\beta_{rD}^3}{a_D}e^{-2b_D} - 2H_D\sigma_D^2 + \frac{8}{3}H_D^2\sigma_D\beta_{rD}^2 + \frac{3}{H_D}\dot{H}^2\tilde{\Omega}_k)_D \right]
\eea
where enriched with the corrections induced from dipole cosmology on baryon asymmetry. Given that during radiation: $H_D \sim \frac{T_D^2}{M_{\text{Pl}}}$, $a_D \sim \frac{1}{T_D}$, thus each term scales as
\bea
&&-12H^4\beta_r^2 \sim -\beta_{rD}^2 T_D^8/M_{\text{Pl}}^4, \quad
-8H^3\beta_r\sigma \sim -\beta_{rD}\sigma_D T_D^6/M_{\text{Pl}}^3,  \\ \nonumber
&&-8H^3\beta_r^3 e^{-2b}/a \sim -\beta_{rD}^3 T_D^7/M_{\text{Pl}}^3, \quad  
-2H\sigma^2 \sim -\sigma_D^2 T_D^2/M_{\text{Pl}},  \\ \nonumber &&\frac{8}{3}H^2\sigma\beta_r^2 \sim +\beta_{rD}^2\sigma_D T_D^4/M_{\text{Pl}}^2
\eea
By taking the dominant term for small $\beta_r$ of $\dot{R}$, i.e., 	$\dot{R} \approx -12H_D^4 \beta_{rD}^2$ in the account of the baryon asymmetry (\ref{dipol}), we have
\be \label{eta}
\eta_B^{\text{dipole}} \approx - \frac{15g_b}{4\pi^2 g_*} \frac{12H_D^4\beta_{rD}^2}{M_*^2 T_D} 
\ee
By requiring $|\eta_B^{\text{dipole}}| \lesssim \eta_B^{\text{obs}} = 8.6 \times 10^{-11}$, for different values of $T_D $ obtain different upper bounds on $\beta_{rD}$, see Table \ref{tab:gb_bounds} and Fig. \ref{fig:Td}.
Note that the subscript $D$ means evaluated at decoupling temperature $T_D$.

Concerning Fig. \ref{fig:Td}, the black curve represents the upper bound \(\beta_{rD}^{\text{max}}\) derived from the GB mechanism, following the scaling relation \(\beta_{\text{max}} \propto T_D^{-7/2}\) \footnote{The strong temperature dependence of the GB bound arises from the scaling of the Hubble parameter during the radiation-dominated era, \(H_D \propto T_D^2 / M_{\text{Pl}}\). Substituting into Eq.~(\ref{eta}) yields \(\eta_B \propto H_D^4 \beta^2 / T_D \propto T_D^7 \beta^2 / M_{\text{Pl}}^6\). For fixed observed baryon asymmetry \(\eta_B^{\text{obs}}\), this implies \(\beta_{\text{max}} \propto T_D^{-7/2}\), explaining the steep decline of the blue curve in Fig. \ref{fig:Td}.}. The green shaded region below the curve indicates parameter space allowed by GB, where the generated baryon asymmetry satisfies \(|\eta_B^{\text{dipole}}| \lesssim \eta_B^{\text{obs}} = 8.6 \times 10^{-11}\). The red shaded region above the curve is excluded, as it would produce an excess of baryon asymmetry inconsistent with observations. Dashed vertical lines mark specific decoupling temperatures \(T_D = 10^{12}\) GeV and \(T_D = 10^{15}\) GeV, while the horizontal dashed line at \(\beta = 1\) indicates the boundary between sub-relativistic and relativistic bulk flows.

At lower decoupling temperatures, such as \(T_D = 10^{12}\) GeV, the GB constraint permits relatively large tilt values up to \(\beta_{rD} \sim 10^7\), corresponding to a bulk velocity \(v_r/c = \tanh \beta_{rD} \approx 1\). However, as \(T_D\) increases, the bound tightens dramatically. At \(T_D = 10^{15}\) GeV, the maximum allowed tilt is only \(\beta_{rD} \approx 1.7 \times 10^{-3}\), corresponding to a non-relativistic bulk velocity \(v_r/c \approx 1.7 \times 10^{-3}\). At \(T_D = 10^{16}\) GeV, the constraint becomes extremely stringent, \(\beta_{rD} \lesssim 5.3 \times 10^{-7}\).

\begin{figure}[htbp]
	\centering
	\includegraphics[width=0.8\textwidth]{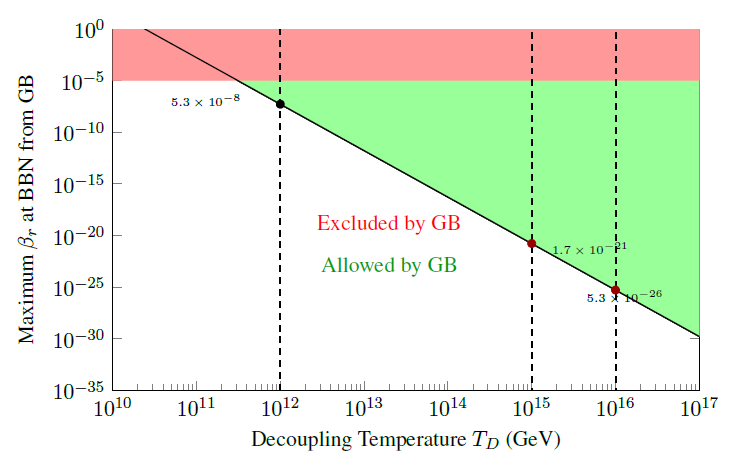}
	\caption{he allowed and excluded regions in the $T_D-\beta_{rD}^{BBN,\max}$ parameter space in light of constraint $|\eta_B^{\text{dipole}}| \lesssim \eta_B^{\text{obs}}$.}
	\label{fig:Tdd}
\end{figure}


\begin{table}[h!]
	\centering
	\caption{Upper bounds on $\beta_r$ from GB for different decoupling temperatures $T_D$}
	\begin{tabular}{lccc}
		\toprule
		\hline
		$T_D$ (GeV) & $\beta_{rD}^{\max}$ (at $T_D$) & $\beta_{r}^{\text{BBN},\max}$ & Consistency with BBN \\
		\hline
		\midrule
		$10^{12}$ & $5.3 \times 10^{7}$  & $5.3 \times 10^{-8}$  & $\checkmark$ \\
		$10^{13}$ & $1.7 \times 10^{4}$  & $1.7 \times 10^{-12}$ & $\checkmark$ \\
		$10^{14}$ & $5.3$                 & $5.3 \times 10^{-17}$ & $\checkmark$ \\
		$10^{15}$ & $1.7 \times 10^{-3}$ & $1.7 \times 10^{-21}$ & $\checkmark$ \\
		$10^{16}$ & $5.3 \times 10^{-7}$ & $5.3 \times 10^{-26}$ & $\checkmark$ \\
		\bottomrule
		\hline
	\end{tabular}
	\label{tab:gb_bounds}
\end{table}
From tilt evolution equation (\ref{beta_eq}) for radiation,  we have
\bea
\dot{\beta}_r \approx -\frac{2}{3}\sigma\beta_r - \frac{2}{3}\frac{\beta_r^2}{a}e^{-2b}~,
\eea
since $3\coth\beta_r - \tanh\beta_r \approx 3/\beta_r - \beta_r \approx 3/\beta_r$. If shear $\sigma$ is small, the second term dominates:
\be
\dot{\beta}_r \approx -\frac{2}{3}\frac{\beta_r^2}{a}e^{-2b}.
\ee
By assuming that $e^{-2b} \sim$ constant, then since $a \propto t^{1/2}$ in radiation \footnote{The relation \(a \propto t^{1/2}\) arises from the Friedmann equation for a radiation-dominated universe with equation of state \(p = \rho/3\) \cite{Trodden:2004st}, and it remains valid for dipole cosmology during the BBN epoch, when radiation is the dominant component. In tilted Bianchi cosmologies, this scaling is consistent with the findings of Ref.~\cite{Lim:2006fk} for radiation-filled models, where the radiation fluid is characterized by an equation of state parameter \(\gamma = 4/3\), and the expansion dynamics follow the standard FLRW scaling at leading order in the tilt. In this context, the derived constraints require \(|\beta_r| \ll 1\), ensuring that the leading-order behavior is accurately described by the conventional radiation-dominated scaling. We also note that alternative parametrizations of the scale factor, such as \(a(t) \sim \sinh(t)^{2/3}\) \cite{Sazhin:2011ze}, are appropriate for the matter-dominated era, reducing to the standard relation \(a(t) \propto t^{2/3}\) at early times. However, such forms are not suitable for the radiation-dominated era relevant to BBN, where the correct scaling is \(a(t) \propto t^{1/2}\).}, as a result
\be
\frac{d\beta_r}{\beta_r^2} \propto -t^{-1/2} dt~,
\ee
meaning that $\beta_r \propto t^{-1/2} \propto T$. So indeed $\beta_r \propto T$, i.e., GB constraint is
\be
\beta_{r}^{\text{BBN}} = \beta_{rD} \times \frac{T_{\text{BBN}}}{T_D}.
\ee
Now by setting $T_{\text{BBN}} = 10^{-3}$ GeV, and different values of $T_D$, one can extract some tight upper bounds for $\beta_{r}^{\text{BBN}}$, as released in Table \ref{tab:gb_bounds} (see also Fig. \ref{fig:Tdd}). 

The black curve represents the upper bound \(\beta_r^{\text{BBN,max}}\) as a function of the decoupling temperature \(T_D\). The green shaded region below the curve is allowed by GB, meaning any \(\beta_r\) value in this region produces a baryon asymmetry consistent with observations (\(|\eta_B| \leq \eta_B^{\text{obs}}\)). The red shaded region above the curve is excluded by GB, as it would overproduce the baryon asymmetry.

For decoupling temperatures \(T_D \gtrsim 10^{12}\) GeV, GB imposes extremely tight constraints on the radiation tilt at BBN, typically \(\beta_r^{\text{BBN}} \lesssim 10^{-8}\) or smaller. These bounds are orders of magnitude stronger than the direct BBN constraint (\(\beta_r < 0.03\)), demonstrating that GB is a highly sensitive probe of early-Universe anisotropy. The scaling \(\beta_{\max}^{\text{BBN}} \propto T_D^{-9/2}\) reflects the combined effect of the GB mechanism (\(\propto T_D^{-7/2}\)) and the tilt evolution (\(\propto T_D^{-1}\)). If GB is the correct mechanism for baryogenesis, the radiation tilt at BBN must have been virtually zero for high decoupling temperatures, implying a remarkably isotropic early Universe.


\section{Conclusion} \label{co}
We have performed a detailed analysis of Big Bang Nucleosynthesis (BBN) and Gravitational Baryogenesis (GB) within the tilted dipole $\Lambda$CDM cosmological framework. This model extends the standard cosmology by introducing an anisotropic metric and separate bulk velocities for matter and radiation, parameterized by the tilt parameters $\beta_m$ and $\beta_r$. Our focus has been on the radiation sector during the BBN epoch ($T \sim 0.1$--$1$ MeV), where the cosmic dynamics are dominated by relativistic species.

The modified expansion rate $H(T)$ in dipole cosmology, derived from the anisotropic field equations, introduces a dependence on the radiation tilt $\beta_r$ and the associated shear $\sigma$. We expressed this modification in terms of an effective change in the number of relativistic degrees of freedom, $\Delta N_{\mathrm{eff}}$, and derived explicit scaling relations for the freeze-out temperatures and abundances of light elements ($^4$He, D, $^3$He, $^7$Li) as functions of $\beta_r$ for both maximal and minimal shear configurations.

The primordial abundances of Helium-4 ($Y_p$) and deuterium (D/H) provide the most stringent and reliable constraints. $Y_p$ is highly sensitive to the neutron-to-proton freeze-out ratio, which is directly affected by changes in $H(T)$. Deuterium abundance, determined by the $D(p,\gamma)^3$He reaction freeze-out, offers complementary sensitivity. A combined $\chi^2$ analysis yields the tightest $2\sigma$ upper limits: $|\beta_r| \lesssim 0.03$ for the maximal shear case and $|\beta_r| \lesssim 0.2$ for the minimal shear case. For comparison, constraints from $\Delta N_{\mathrm{eff}}$ alone give $|\beta_r| \lesssim 0.047$ (maximal shear) and $|\beta_r| \lesssim 0.221$ (minimal shear), demonstrating that the combined $\chi^2$ analysis from individual element abundances provides more stringent constraints. Both sets of bounds are upper limits; $\beta_r = 0$ is always allowed by the BBN constraints. The lithium-7 problem---the persistent discrepancy between standard BBN predictions and observations---remains unsolved in this framework, as the $\beta_r$ values required to alleviate it ($|\beta_r| \sim 1$) are conclusively excluded by the helium-4 constraints.

We further investigated implications for gravitational baryogenesis, a mechanism that generates the baryon asymmetry through a CPT-violating coupling $(\partial_\mu R) J_B^\mu$. Computing the Ricci scalar $R$ for the dipole metric, we found leading-order corrections proportional to $H^4\beta_r^2$. The observed baryon asymmetry $\eta_B^{\rm obs} \approx 8.6 \times 10^{-11}$ then imposes upper limits on $\beta_r$ at the GB decoupling temperature $T_D$. Under the reasonable assumption that $\beta_r \propto T$ during radiation domination (as motivated by the tilt evolution equation), these limits translate to extremely stringent constraints on $\beta_r$ at the BBN epoch: $|\beta_r^{\rm BBN}| \lesssim 10^{-8}$ for $T_D \gtrsim 10^{12}$ GeV. These GB bounds are orders of magnitude tighter than those from direct BBN, indicating that if GB operates at high scales, the radiation tilt must have been exponentially suppressed in the early-Universe.

Our results carry several key implications for the tilted dipole cosmology and related anisotropic models. First, any large-scale bulk flow of radiation during the BBN era must have been very small, with a corresponding velocity $v_r/c = \tanh\beta_r \lesssim 0.03$ in the maximal shear case (the tightest constraint) and $v_r/c \lesssim 0.2$ in the minimal shear case (the weakest constraint). This indicates that a high degree of isotropy was already established by $T \sim 1$ MeV. Second, the strong BBN and GB constraints are consistent with the model's late-time attractor behavior, where shear decays as $\sigma \propto e^{-3Ht}$, ultimately leading to the observed isotropy of the CMB. The tilt parameter $\beta_r$ must have decreased significantly from any possible high-redshift value to meet the tight bounds at BBN. Third, the tilted dipole $\Lambda$CDM model remains viable, but only in a parameter region where early-Universe anisotropy is heavily suppressed; it cannot resolve the lithium abundance anomaly without violating other well-measured elemental abundances. Fourth, primordial light elements, especially $^4$He and deuterium, serve as powerful probes of pre-CMB anisotropy. Future precision measurements of these abundances, along with improved constraints on $\Delta N_{\mathrm{eff}}$ from next-generation CMB experiments, will further test anisotropic cosmological scenarios and narrow the allowed parameter space for early-Universe departures from isotropy.

In summary, while the tilted dipole cosmology provides a theoretically motivated framework for incorporating observed large-scale bulk flows and CMB dipoles, it must conform to stringent limits on early-Universe anisotropy set by primordial nucleosynthesis and baryogenesis. The model survives only if the radiation tilt is kept very small ($|\beta_r| \lesssim 0.03$) during BBN, pointing to a remarkably isotropic universe from at least the first few minutes onward.
	
\begin{acknowledgments}
The author sincerely thanks M.M. Sheikh-Jabbari for technical discussions and insightful comments on the manuscript.
\end{acknowledgments}

\appendix
\section{Derivation of the sum rule in dipole cosmology}\label{A}	

In this appendix, we provide a detailed derivation of the sum rule
\begin{equation}\label{A1}
	\tilde{\Omega}_{r} + \tilde{\Omega}_{k} + \tilde{\Omega}_{\sigma} = 1, 
\end{equation}
which plays a central role. This derivation clarifies that the sum rule is not an externally imposed constraint but rather a fundamental consistency condition following directly from the Einstein field equations for the tilted Bianchi metric.
\\
We begin with the modified Friedmann equation for the dipole cosmology metric during radiation domination, as given in Eq.~(\ref{Hubble_eq}). 
This equation follows from the Einstein equations $G_{\mu\nu} = 8\pi G T_{\mu\nu}$ applied to the tilted Bianchi metric (\ref{metric}). The three terms on the right-hand side of the Eq.~(\ref{Hubble_eq})  encapsulate distinct physical contributions to the overall dynamical evolution: first, the spatial curvature term, which is governed by the parameter \(A_0\) and reflects the geometry of the universe; second, the radiation energy density term, which is modulated by the tilt parameter \(\beta_r\) to account for directional dependencies or preferred-frame effects in the radiation fluid; and third, the shear anisotropy term, given by \(\sigma^2/9\), which quantifies the contribution from anisotropic expansion or distortion in the spacetime fabric.
\\
Following Ref.~\cite{Ebrahimian:2023svi}, we define the following dimensionless density parameters:
\bea
&&	\tilde{\Omega}_{r} \equiv \frac{\rho_{r}(1 + \frac{4}{3}\sinh^{2}\beta_{r})}{3H^{2}},  \label{A3} \\
&&	\tilde{\Omega}_{k} \equiv \frac{A_{0}^{2}}{a^{2}e^{4b}H^{2}},  \label{A4} \\
&&	\tilde{\Omega}_{\sigma} \equiv \frac{\sigma^{2}}{9H^{2}}. \label{A5}
\eea
These definitions are chosen so that each parameter represents the fraction of the total energy density contributed by a specific component: radiation (with tilt modifications), spatial curvature, and shear, respectively.
\\
Dividing the Friedmann equation (\ref{Hubble_eq}) by $H^2$ on both sides yields:
\begin{equation}\label{A6}
	1 = \frac{A_{0}^{2}}{a^{2}e^{4b}H^2} + \frac{\rho_{r}(1 + \frac{4}{3}\sinh^{2}\beta_{r})}{3H^2} + \frac{\sigma^{2}}{9H^2}. 
\end{equation}
Substituting the definitions from Eqs.~(\ref{A3})--(\ref{A5}) into Eq.~(\ref{A6}), we immediately obtain the sum rule (\ref{A1}), meaning that 
it is a direct consequence of the Friedmann equation, not an additional constraint imposed by hand.

\end{document}